\pdfoutput=1 
\documentclass{JINST}

\usepackage[subrefformat=parens,labelformat=parens]{subfig}
\usepackage{lineno}

\title{A Noble Gas Detector with Electroluminescence Readout based on an Array of APDs}

\author{
B. Bourguille$^a$, 
A. Garcia Soto$^a$, 
I. Gil-Botella$^b$, 
T. Lux$^a$\thanks{Corresponding author.}, 
C. Palomares$^b$, 
F. Sanchez$^a$~
and R. Santorelli$^b$\\
\llap{$^a$}Institut de F\'{i}sica d\'{}Altes Energies (IFAE), The Barcelona Institute of Science and Technology, Campus UAB,  Bellaterra (Barcelona), Spain\\
\llap{$^b$}Centro de Investigaciones Energ\'{e}ticas, Medioambientales y Tecnol\'{o}gias (CIEMAT), Madrid, Spain\\
E-mail: \email{Thorsten.Lux@ifae.es}}

\abstract{We present the results of the operation of an array of avalanche photodiodes (APDs) for the readout of an electroluminescence detector. The detector contains 24 APDs with a pitch of 15 mm between them allowing energy and position measurements simultaneously. Measurements were performed in xenon (3.8 bar) and argon (4.8 bar) showing a good energy resolution of 5.3\% FWHM at 60 keV in xenon and 9.4\% in argon respectively. Following Monte Carlo studies the performance could be improved significantly by reducing the pitch between the sensors.
}


\begin{document}

\section{Introduction}
Electroluminescence (EL) detectors were and are in a wide range of experiments as for example to detect low energy X-rays in the satellite experiment BeppoSax \cite{manzo1997high} or for dark matter \cite{Aprile:2005ww} and double beta decay experiments \cite{Granena:2009it}.\\
An EL detector is a gas detector in which the more commonly used charge amplification is replaced by a region with an electric field high enough that electrons released by an ionization are accelerated to energies to excite the gas molecules without ionizing them. In the de-excitation process photons are emitted isotropically. This light is referred to as secondary scintillation or EL light. 
This EL region is normally realized by installing two parallel wire meshes, called EL meshes in the following, with a gap of a few mm between them. The wavelength of the emitted EL photons depends on the gas used in the detector. For argon the light has mainly a wavelength of 128 nm and for xenon 172 nm. The intensity of the EL light is directly proportional to the energy deposited in the detector. With an array of suitable light sensors placed behind the two EL meshes it is possible not only to measure the total energy deposited in the detector but to reconstruct the ionization event. \\
An advantage of EL detectors is that they provide a high gain, number of EL photons per primary electron, of up to a few thousands. The light gain, $Y$, depends on the applied EL field, $E$ [kV/cm], the gap, $d$ [cm], between the EL meshes and the operation pressure, $p$ [bar], and can be described by an empirical equation which depends on the used noble gas \cite{Monteiro:2007vz} 
\cite{Monteiro2008167}:
\begin{eqnarray}
Y_{Xe}&=&\bigg(140\frac{E}{p}-116\bigg)\cdot p \cdot d	\\
Y_{Ar}&=&\bigg(81\frac{E}{p}-47\bigg)\cdot p \cdot d		
\end{eqnarray}
From these equations one can easily deduce that the gain increases with increasing pressure, a high pressure might be desired for applications which require a detector with high stopping power. In contrast to this the achievable charge gain drops for argon and xenon with increasing pressure. The high gain allows to detect charge depositions of below 1 keV with EL detectors. Another feature of EL detectors is the excellent energy resolution they provide. While the exponential nature of the charge amplification introduces large gain fluctuations especially at limited gains, this contribution is drastically reduced in EL detectors. In the case of a high detection efficiency for the photons this, in combination with the high gain, leads to results of the energy resolution close to the intrinsic one given by the fluctuations in the number of primary electrons, $N_e$: $\sigma_{Ne}=\sqrt(FW/E_X)$ with $F$ being the Fano factor, $W$ the average energy to produce an electron-ion pair and $E_X$ the energy of the X-ray.\\
The choice of the light sensors for the detection of the EL light is of great importance. Here, we present a study using Avalanche PhotoDiodes (APDs) for this purpose. These solid state devices have various advantages, they are resistant to very high pressures, can be produced in sizes interesting for imaging applications and they are not only directly sensitive to the Xe light but also to the Ar EL light while in the case of PMTs wavelength shifter are normally used.

\section{Experimental Setup} \label{sec_setup}
\begin{figure}[h!!]
\centering
\subfloat[]{\label{fig_setup_a} \includegraphics[width=0.6\textwidth]{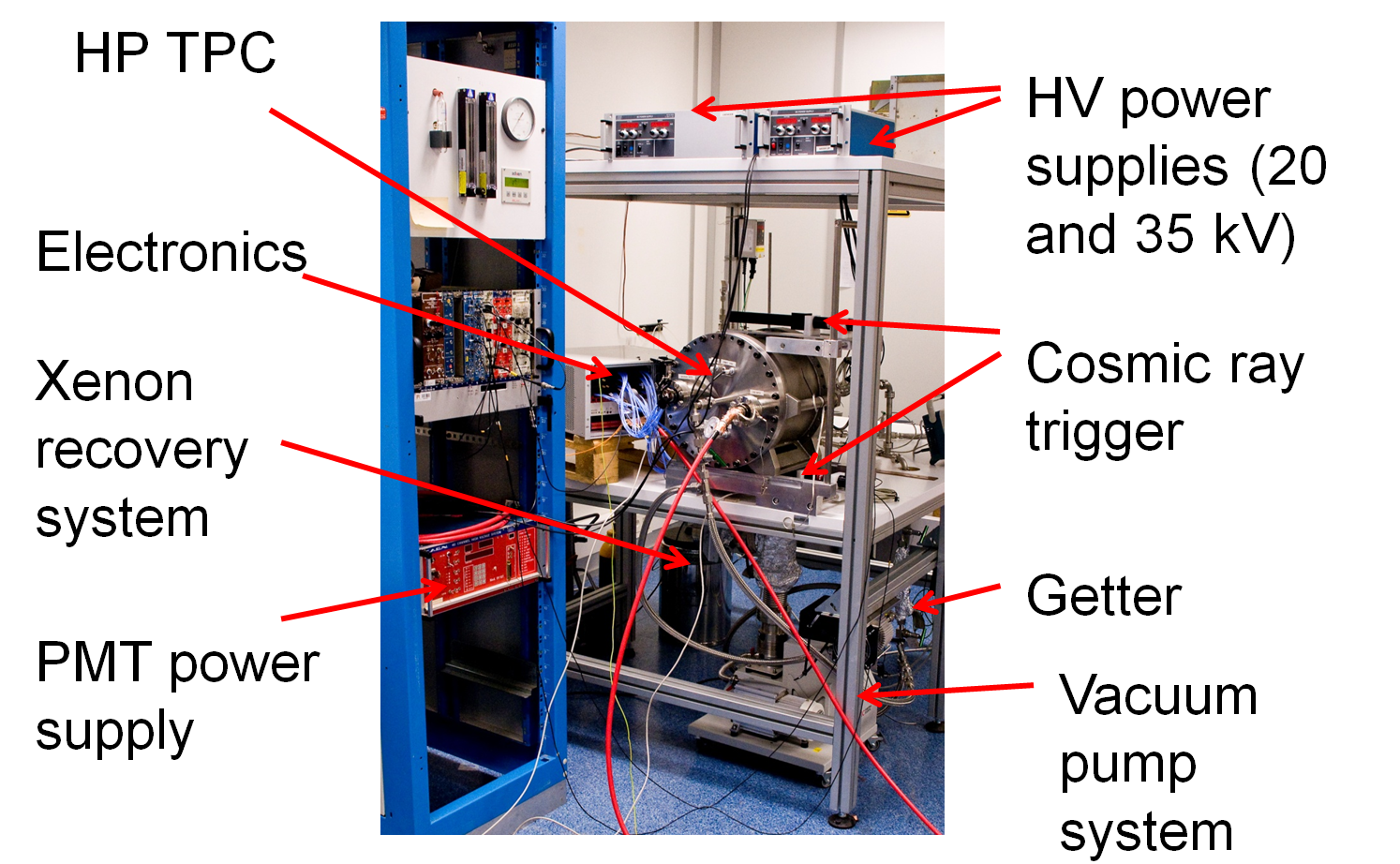}}
\hspace{0.001\linewidth}
\subfloat[]{\label{fig_setup_b} \includegraphics[width=0.35\textwidth]{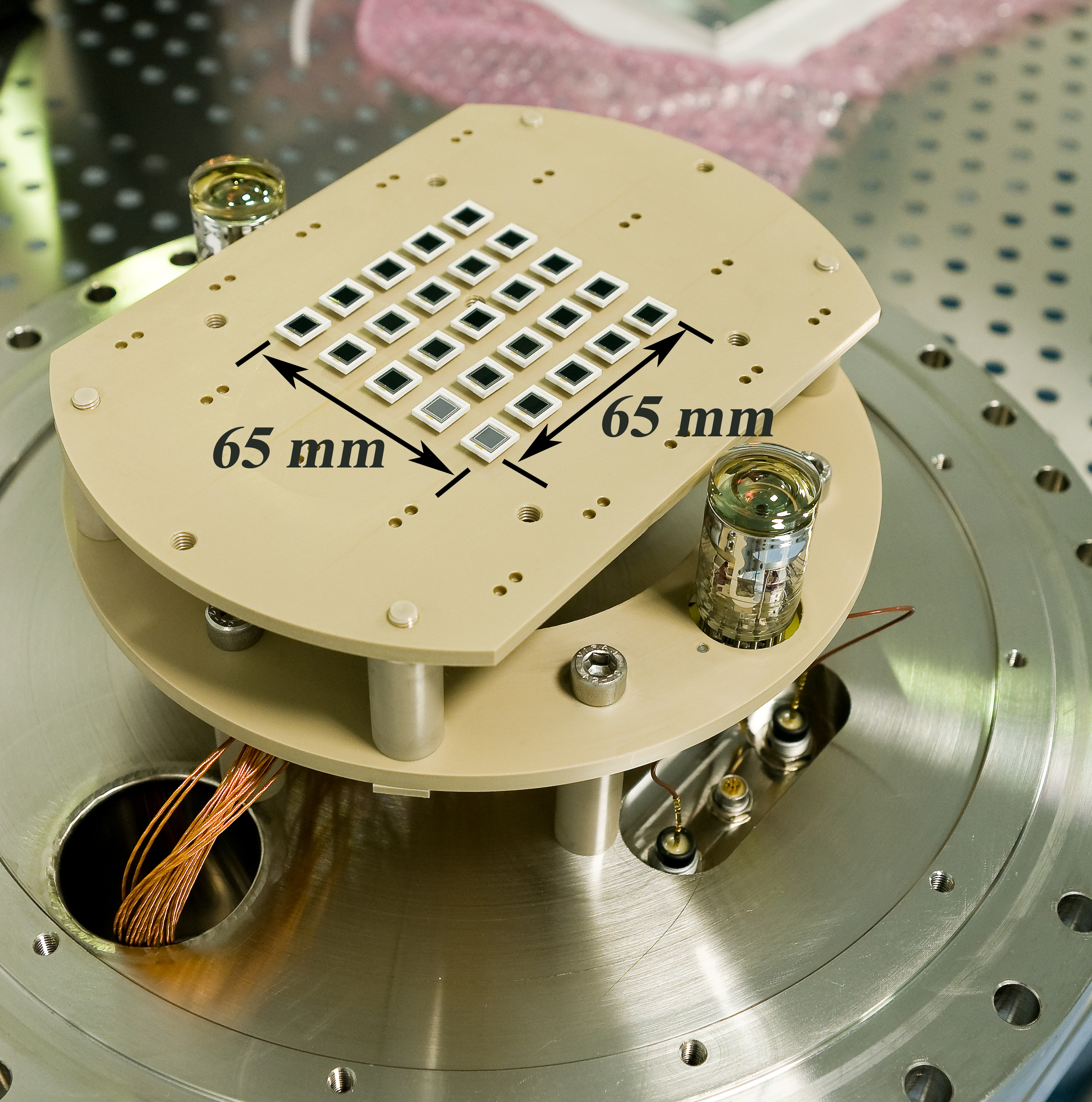}}
\caption{(a) The full setup used for the studies. (b) The readout plane with 25 APDs and the two PMTs. }
\label{fig_setup}
\end{figure}

The apparatus is described in detail somewhere else \cite{1748-0221-10-03-P03008}. Here, we only can give a short summary of it: 
The X-ray conversion volume has a diameter of about 20 cm and a drift distance of about 11 cm. However, only the central region of 6x6 cm$^2$ is equipped with 25 APD sensors (Hamamatsu S8664-SPL) of which 24 are read out (Fig.~\ref{fig_setup}). The APDs have an active area of about 5x5 mm$^2$ and are arranged in the $xy$ plane as a 5x5 matrix with a pitch of 15 mm between the APDs ( $\approx$ 11\% area coverage). The APDs are standard devices from  which have been made sensitive to VUV photons by removing the protective layer covering the sensitive area.  They have a high quantum efficiency for both, xenon and argon light and were characterized in a previous study \cite{Lux201211}. The EL amplification region is terminated by two parallel meshes with a gap of 7 mm between them. 
A $^{241}$Am source ($\approx$5~kBq) was used which provides several of gammas lines, the most significant ones being at 13.95, 17.75, 20.78, 26.3 and 59.5 keV \cite{Lepy2008715,Gunnink1976}. In xenon several additional escapes peaks in the range between 26 and 34 keV can be seen due to {\it{K}} shell X-ray fluorescence lines (29.46, 29.78, 33.59 and 34.42 keV \cite{507079}), with the most pronounced one around 30 keV.\\

\section{Event Reconstruction}
The event reconstruction is based on three main steps: first, extracting the signal, amount of light detected, from each APD in the sensor plane; second, applying an intercalibration to equalize the APD responses; third, determining the energy deposited in the detector and the 2D reconstruction of the position of the interaction position. In this section the principle of the event reconstruction will be summarized, for a mathematically detailed description the interested reader is referred to \cite{1748-0221-10-03-P03008}.   

\subsection{Signal Extraction}
\begin{figure}[h!!]
\centering
\includegraphics[width=0.4\textwidth]{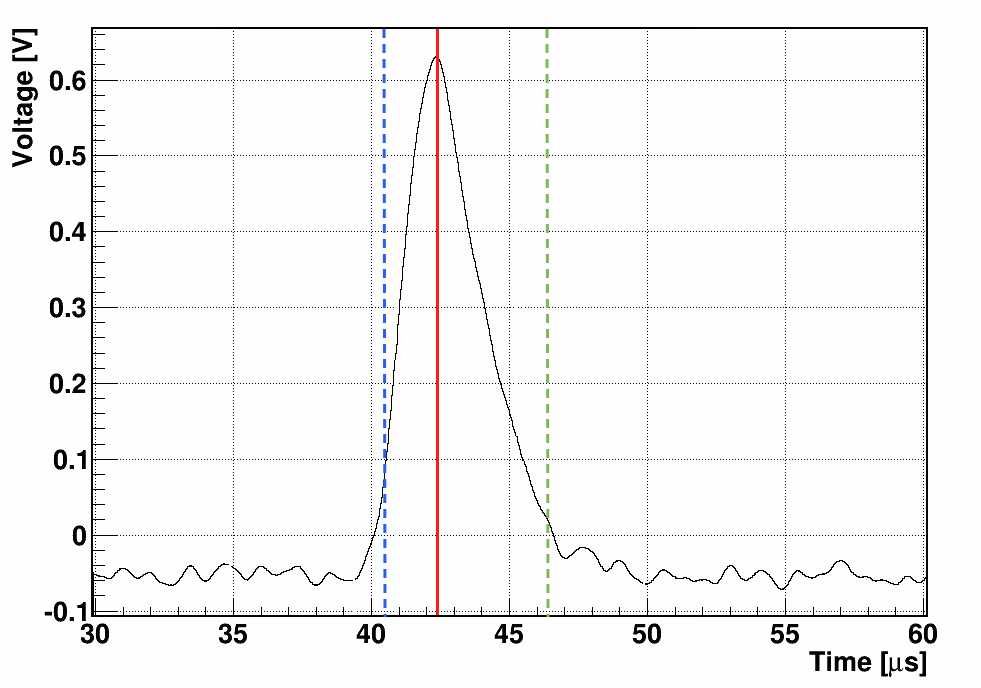}
\caption{Typical waveform from the APD with highest amplitude. The solid red line defines the time of the maximum. The dashed lines indicate the times $t_{up}$ (green) and $t_{low}$ (blue) chosen for the integration window to optimize the energy resolution. }
\label{waveform}
\end{figure}

In this step the amount of light detected per APD, $E_i$, is extracted from the waveforms which are recorded by the ADC for every APD independently. An example of a waveform is shown in Fig.~\ref{waveform}. First, the APD with the maximal amplitude is identified and the position of the maximum of the pulse is determined. Afterwards the waveform is integrated for all waveforms within a time window around the position of the maximum. The limits of the time window were optimized based on the energy resolution achieved. Then the obtained integrals are divided by the width of the time window in bins. Finally these values are corrected for the pedestals which are calculated with the same integration method using a software trigger. 

\subsection{APD Intercalibration} \label{sec_cal}
\begin{figure}[ht!]
\centering
\subfloat[]{\label{fig_cal_a} \includegraphics[width=0.3\textwidth]{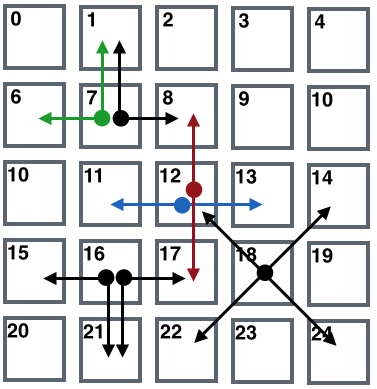}}
\hspace{0.001\linewidth}
\subfloat[]{\label{fig_cal_b} \includegraphics[width=0.5\textwidth]{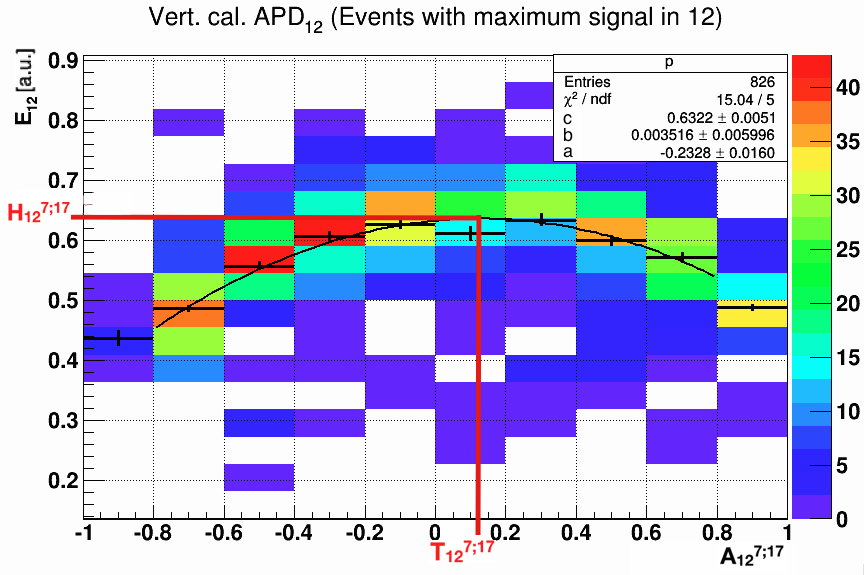}}
\caption{(a) Schematic view of the APD plane showing examples of the 8 possible asymmetry couplings. (b) Signal in the central APD versus the asymmetry between the signal in the side APDs. Both $T$ and $H$ are used to determine the intercalibration factors.}
\label{fig_cal}
\end{figure}

In the next step the signal from each APD was corrected for the gain differences between the individual APDs. Since due to its complexity the installation of a LED-based calibration system was not a suitable solution, an intercalibration method was developed based on the sensor response to the EL light from 59.5 keV X-rays. This method relies on two basic assumptions: 1) the EL light distribution on the sensor plane is radial symmetric and 2) the amount of EL light produced does not depend on the position. In this case there are for each APD 8 combinations of neighbor APDs (see Fig.~\ref{fig_cal_a}) which can be intercalibrated. The method is based on the definition of two variables: 1) {\it Asymmetry}, $A_i^{j;k}=\frac{E_{j}-E_{k}}{E_{j}+E_{k}}$, which computes the light sharing between APD$_j$ and APD$_k$ when the maximum integrated signal is at APD$_i$ with the constraint that the APDs$_{j,k}$ are at identical distance to APD$_i$; 2){\it Maximum Energy} measured in the central APD$_i$.\\
Plotting {\it Asymmetry} versus the signal measured in the central APD one obtains a distribution which follows a second order polynomial with the maximal signal in the central APD at asymmetry 0 in the case of identical gains in the neighboring APDs. Any different position of the maximum,$T_i^{j;k}$, is the consequence of a gain difference and can be used to calculate an intercalibration factor for the two side APDs. This procedure is performed using every APD in the readout plane as central APD. As an extra handle also the {\it Maximum Energy} ($H_i^{j;k}$; see Fig.~\ref{fig_cal_b}) can be determined from this plot and can also be used to intercalibrate the different APDs. With identical gain of all APDs, all $H_i^{j;k}$ are supposed to have the same value. \\
The variables $T_i^{j;k}$ and $H_i^{j;k}$ are used to define a $\chi^2$ function which minimization provides the final calibration factors, $c_i$. This process can be performed iteratively but as we showed in \cite{1748-0221-10-03-P03008} the result converges already after the first iteration towards the final values. \\
This intercalibration method requires that there is enough light detected by the side APDs to calculate reliably the {\it Asymmetry} taking into account the noise of the APDs. While this condition is fulfilled for the Xe measurements, the lower light yield in Ar did not allow for the setup used in this measurements the determination of the intercalibration factors. 
    
\subsection{Energy and Position Reconstruction}
The energy, $Q$, of the X-ray and the point position of the conversion in the $x,y$ projection of the event are reconstructed comparing the measured light distribution, the signals measured by the different APDs, with an expected one, the so-called light-profile ($F_{ij}$), using a maximum likelihood fit. The performance of the reconstruction method depends strongly on the correctness of the used light profile. We studied two different methods to determine the light-profile, an iterative method which relies on the measured APD signals and was developed for a dual-phase xenon scintillation detector \cite{Solovov201212} and an analytic method in which the light-profile is obtained from a MC program which takes into account all relevant geometric and physical effects as e.g. the mesh transparency, diffusion, light yield, etc.. Both methods deliver quite similar light profile distributions. The reconstruction method allows to estimate position and energy of an event minimizing the following expression:
\begin{equation}\label{eq_chi2}
\chi^2=\sum_{i=0}^{\#apds}\sum_{j=0}^{\#apds}\left(E_i-QF_i(x,y)\right)\Sigma_{ij}^{-1}\left(E_j-QF_j(x,y)\right)
\end{equation}
Here, $\Sigma_{ij}$ is the covariance matrix which takes into account the correlation between the pedestals and primary electron fluctuations from the different APDs.  
The performance of both methods is similar at low light gains while at high light gains the analytic method provides a by about 10\% better energy resolution.   

\section{Results}
We commissioned the system using xenon as filling gas and an intensive study of the detector performance was carried out afterwards with this gas. After the measurements with Xe we filled the detector with Ar. In both cases we performed scans of the various operation parameters as drift field, EL field and APD bias voltage. Due to the limited space, we concentrate mainly on the argon results and present only shortly the results from the Xe study and suggest the interested reader to read \cite{1748-0221-10-03-P03008} for more details about the Xe study. 
 
\subsection{Xenon Results}
\begin{figure}[ht!]
\centering
\includegraphics[width=0.5\textwidth]{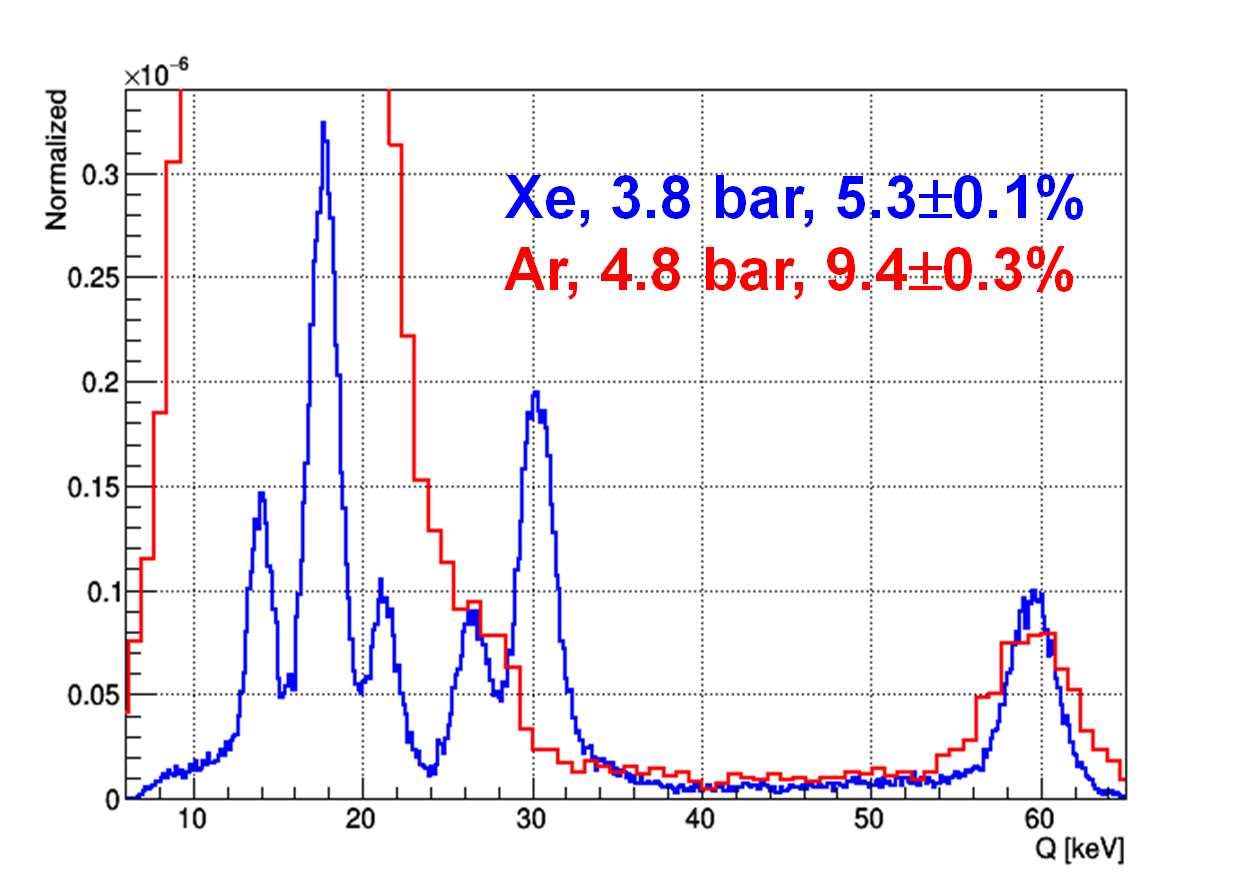}
\caption{Spectra of $^{241}$Am measured in Xe (blue) and Ar (red). The detection threshold for both gases is below 13 keV. The 26 and 30 keV lines which can be observed in Xe but not in Ar have their origin in X-ray fluorescence with Xe atoms. }
\label{fig_enres}
\end{figure}

The presented Xe data was taken at a pressure of 3.8 bar and processed in the above described way. We achieved an energy resolution of 5.3\% at 59.5 keV (Fig.~\ref{fig_enres}). This result is significantly worse than the theoretical limit of about 1.7\% despite the high light gain of about 1100 but this discrepancy can be explained by the limited sensor coverage, about 11\%, in the readout plane. It was shown with a MC which reproduced well our results, 5.2\% instead of the measured 5.3\%, that with a higher sensor coverage the energy could be reduced to about 2.2\% (96\% sensor coverage) or 3\% (25\% sensor coverage). We used the setup also to determine the achievable point resolution of this setup using cosmic muon data. For vertical tracks point resolutions of about 0.5 mm were achieved \cite{1748-0221-10-03-P03008}. 

\subsection{Argon Results}
After the measurements with xenon the detector was filled with 4.8 bar of argon, we chose a higher pressure than in Xe to compensate the lower light gain in argon. The study the direct sensor response to Ar light is of special interest since few sensors are directly sensitive to this wavelength. The common approach is to use wavelength shifter as tetraphenyl butadiene (TPB) for this application. However, this approach introduces some inconveniences which let it seem interesting to test directly sensitive sensors. 

\subsubsection{Control Plots}
\begin{figure}[ht!]
\centering
\subfloat[]{\label{control_a} \includegraphics[width=0.4\textwidth]{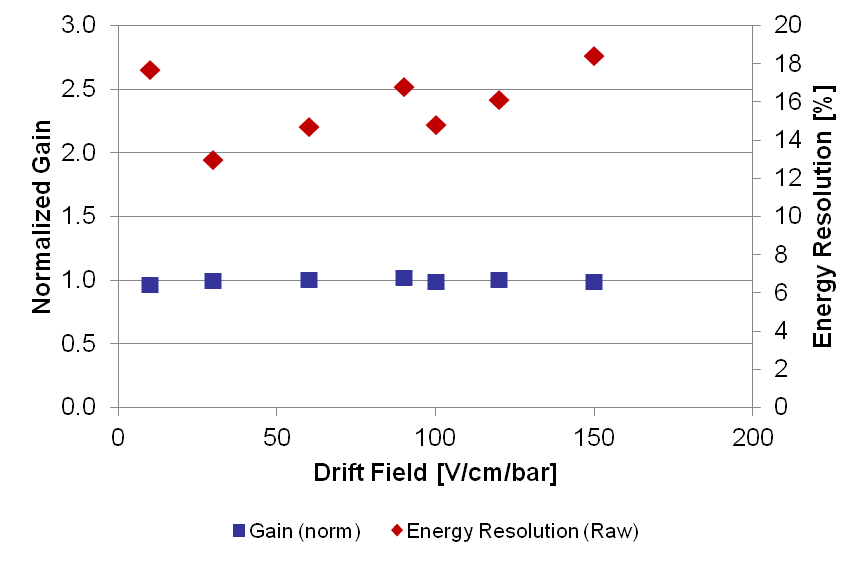}}
\hspace{0.001\linewidth}
\subfloat[]{\label{control_b} \includegraphics[width=0.4\textwidth]{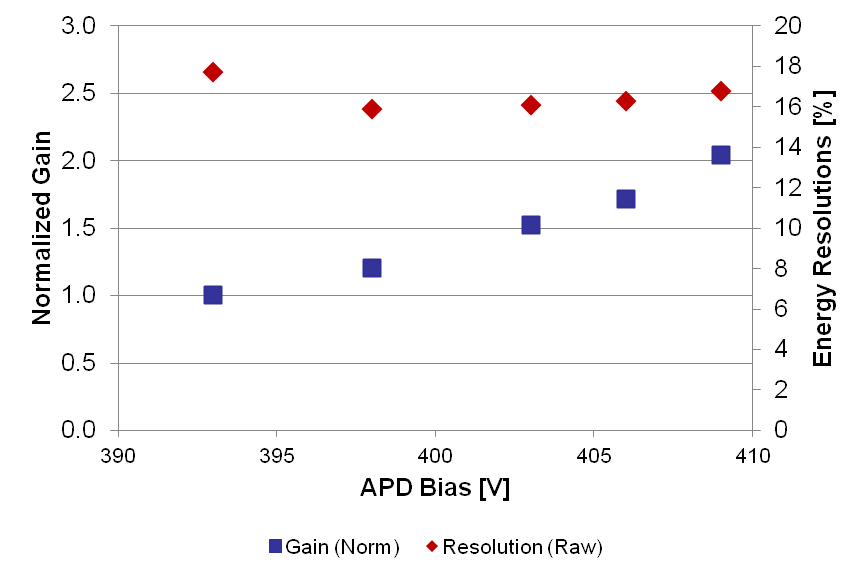}}
\hspace{0.001\linewidth}
\subfloat[]{\label{control_c} \includegraphics[width=0.4\textwidth]{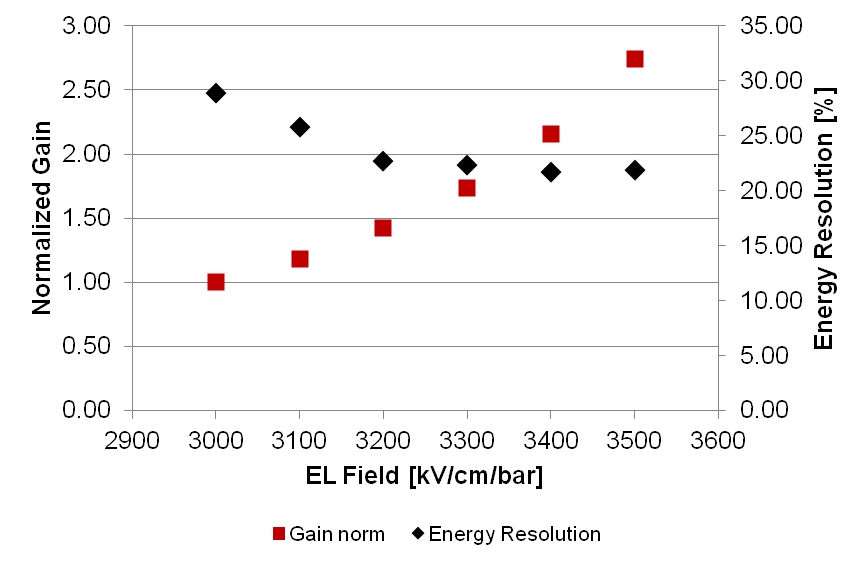}}
\caption{Energy resolution and normalized gain versus (a) drift field (b) APD bias voltage and (c) EL field.  }
\label{controlplots}
\end{figure}

To study the performance of the detector when filled with argon, the EL light intensity and the energy resolution in function of the different operation parameters (APD Bias voltage, EL field and drift field) were measured. These values are determined from the ${^{241}}$Am spectrum (red line in Fig.~\ref{fig_enres}) by fitting a Gaussian to the peak of the 59.5 keV line. During the scans two operation parameters were fixed, while the third was varied. Fig.\ref{controlplots} shows the results these parameter scans. The results follow the tendency for literature results with single sensors. 

\subsubsection{Energy Resolution}
The energy resolution was determined for 59.5 keV, see Fig.~\ref{fig_enres}, to be $9.4\pm0.3\%$ applying a correction for the radial position and a cut on the radius of the reconstructed position of 10 mm. The energy resolution is significantly worse than for xenon and the theoretical limit for argon but this can be explained by the large pitch between the APDs which complicates in Ar the event reconstruction. Since the EL gain in Ar at 4.8 bar is only about 650, it is not possible to apply the intercalibration method described in sec.~\ref{sec_cal}. In addition the high transparency of Ar to 59.5 keV X-rays limits the amount of available statistics for the intercalibration. However, the achieved energy resolution is a promising result and indicates that with a readout with higher sensor density, this result could be significantly improved.  

\section{Conclusions}
We demonstrated that an array of Hamamatsu  APDs S8664-SPL is suitable to detect directly the EL light emitted in xenon and argon. We achieved an energy resolution of 5.3\% (FWHM) in Xe for 59.5 keV and 9.4\%  in Ar. In Xe we achieved point resolutions of about 0.5 mm with cosmics \cite{1748-0221-10-03-P03008}. \\
The obtained results are significant worse than the theoretical ones, $R_{theo}$, which are given by the Fano factor, $F$, the energy
to create a electron-ion pair, $W$, (both depend on the filling as) and the energy of the X-ray, $E_{Xray}$ by $R_{theo}=sqrt{FE_{Xray}/W}$. The theoretical values for xenon and argon are $\approx$ 1.7 and $\approx$ 2.2 \% (FWHM) respectively. We used MC studies to understand this discrepancy and can explain it by the small sensor coverage in the setup used for this study, only about 10\% of the APD plane is sensitive area. The MC reproduces well the energy resolution result for Xe (MC:5.2\%, Data: 5.3\%) and allows to extrapolate to geometries with higher sensor density predicting for this case results similar to the one stated in \cite{Bolozdynya1997225}. For argon a denser packed sensor plane also would give in addition the possibility to apply an intercalibration to the APDs which would help to improve the result further.\\ 
The results with argon are of special interest since they demonstrate that the Hamamatsu 8664-SPL APDs are suitable to detect directly the argon light without the need to use wavelength shifter material. To lower the detection threshold in Ar further, as it might be interesting for dark matter experiments, we consider an interesting option to replace the meshes by a THGEM \cite{1748-0221-7-06-P06012}\cite{Monteiro201218}.

\acknowledgments
The authors acknowledge the support received from the Ministerio de Ciencia e Innovacion under grants FPA 2011-29823-C02-02 and FPA2014-59855-P, Consolider Ingenio Project CSD2008-0037 (CUP), Consolider Ingenio Project CSD-2007-00042 (CPAN) and Centro de Excelencia Severo Ochoa SEV-2012-0234, some of which include ERDF funds from the European Union. 

\bibliography{EL_DB}{}

\bibliographystyle{JHEP}

\end{document}